\documentclass[11pt]{article}
\usepackage[utf8]{inputenc}
\usepackage[T1]{fontenc}
\usepackage[a4paper,margin=1in]{geometry}
\usepackage{lmodern}
\usepackage{microtype}
\usepackage{graphicx}
\usepackage{xcolor}
\usepackage{amsmath,amssymb}
\usepackage{longtable,booktabs,array}
\usepackage{calc}
\usepackage{etoolbox}
\makeatletter
\patchcmd\longtable{\par}{\if@noskipsec\mbox{}\fi\par}{}{}
\makeatother

\providecommand{\real}[1]{#1}
\usepackage{enumitem}
\usepackage{eurosym}
\usepackage[most]{tcolorbox}
\usepackage{caption}
\usepackage{authblk}
\usepackage[hidelinks]{hyperref}
\usepackage[super,comma,sort&compress]{natbib}

\definecolor{boxrule}{HTML}{003E6B}
\definecolor{boxfill}{HTML}{F4F8FB}
\title{\bfseries The Dataset Friction Framework:\\ measuring user-facing friction as a complement to FAIR}
\author[1]{Emma Pidduck}
\author[1]{Umberto Modigliani}
\affil[1]{European Centre for Medium-Range Weather Forecasts (ECMWF), Bonn, Germany}
\date{\today\\[2pt]\small Corresponding author: \href{mailto:emma.pidduck@ecmwf.int}{emma.pidduck@ecmwf.int}}

\begin{document}
\maketitle

\begin{abstract}
Open research data services have matured to the point where the cost of sustaining them at scale has become a primary design constraint, driving providers to make deliberate choices that may reduce user convenience to keep the service viable. The FAIR (Findable, Accessible, Interoperable, Reuseable) principles describe whether a dataset is well stewarded, and FAIR compliance is often treated as a proxy for usability. FAIR does not capture the cost to a user of finding, accessing, interpreting, and applying a dataset. We introduce the Dataset Friction Framework (DFF) as a complement to FAIR, directly addressing usability. DFF measures user-facing friction across six dimensions, distinguishing engineered friction (deliberate data provider design choices that sustain a service) from accidental friction (defects that require remediation). The framework is validated against 18,556 support tickets from the European Centre for Medium-Range Weather Forecasts (January 2024 to May 2026), which serves 280,000 registered users. Restricting the analysis to tickets raised by external reporters reduces the corpus by 12.3\%, but every dimension's internal-staff share falls below this baseline --- confirming that the reported friction signals are genuinely user-facing. We then assess three real datasets across three providers and show that FAIR compliance and DFF friction can disagree in both directions: a 92\% FAIR-compliant dataset can still carry substantial friction, and a 42\% FAIR score can be an artefact of anti-scraping policy rather than poor stewardship. The two measures are non-redundant and jointly informative: FAIR compliance does not predict DFF friction in either direction. This constitutes the first large-scale empirical application of the framework; cross-institutional validation is identified as the immediate next step.
\end{abstract}


\bigskip
\section{Introduction}

Openness, as codified by the FAIR principles \cite{wilkinson2016}, has become a widely accepted baseline for scientific data stewardship across all domains. That is, data should be Findable, Accessible, Interoperable and Reuseable. In the weather and climate community, the transition from restrictive licensing to open access has been among the most significant infrastructural shifts of the past decade, driven by policy instruments such as the World Meteorological Organization's Resolution 1 \cite{wmo2021} and by the longer-standing argument that open access to environmental data is a precondition for addressing problems at global scale \cite{overpeck2011}. Providers have invested heavily in metadata standards, open licences, persistent identifiers, and machine-readable interfaces, and many prominent meteorological datasets now score strongly against FAIR assessment tooling \cite{devaraju2021}.

Two gaps in FAIR's coverage have nevertheless become evident as open data has matured from principle to operational practice. The first is experiential: a dataset can be findable in a catalogue, accessible through a documented interface, interoperable with community tools, and legally reusable under a clear licence, and still require hours of a new user's time before a first usable file is retrieved. The second is structural: sustaining open data services at scale requires design choices --- release latency, resolution caps, registration, tiered delivery, and differentiated support levels --- that reduce user convenience to enable provider sustainability, equity between user classes, or the continued funding of open provision itself. FAIR addresses neither gap directly nor was it designed to do so; what is left unaddressed reflects the stated scope of the framework rather than any defect in FAIR. The operational consequence, however, is a class of questions that service providers encounter daily without a shared vocabulary in which to answer them. The practical stakes are significant: ECMWF's own transition from a licensing model that generated \euro{}123 million over eleven years to full Creative Commons Attribution 4.0 (CC BY 4.0) provision \cite{pidduck2026} illustrates the scale of the sustainability challenge that open data providers face.

This paper introduces the Dataset Friction Framework (DFF) as a complement to FAIR. The shared goal of both is the productive reuse of open data --- to turn a published dataset into as many genuine users as possible; FAIR addresses whether a dataset can, in principle, be found and used, while DFF addresses what it costs a user to do so in practice. The central commitment of the framework is that the deliberate design choices through which providers sustain open services --- engineered friction --- constitute an analytic category, visible and defensible, rather than being treated as defects to be engineered away. The term data friction has a prior history in science and technology studies, originating with Edwards~\cite{edwards2010} and developed by Bates~\cite{bates2018} among others; our contribution is to operationalise that lineage for operational data service providers, with named dimensions, scoring cues, and a design orientation. The framework is then validated empirically against 18,556 user support tickets from a 280,000-user weather and climate data infrastructure, demonstrating that a 92\% FAIR-compliant system nonetheless generates measurable user-facing friction across all six dimensions.

The paper makes four contributions. First, it introduces ``engineered friction'' as a distinct analytic category and presents a menu of recurring sustainability trade-offs that providers use to configure open services. Second, it develops DFF as a complement to FAIR across six dimensions, with scoring cues that a reviewer with working knowledge of a dataset can apply. Third, it provides large-scale empirical validation: an analysis of 18,556 support tickets, two user surveys, and two public software repositories that together establish friction patterns across user populations, channels, and institutional boundaries. Fourth, it demonstrates that FAIR compliance and DFF friction are non-redundant measures, and that their relationship is non-monotonic: a dataset can score highly on one and poorly on the other, and a low FAIR score does not reliably indicate user-facing friction. Three real datasets --- ECMWF's Integrated Forecasting System (IFS) and Artificial Intelligence Forecasting System (AIFS) Open Data, ERA5 via the Copernicus Climate Data Store, and the German national meteorological service's geoportal --- illustrate three distinct FAIR--DFF relationships, and one of these reveals a previously undocumented failure mode of passive FAIR assessment tooling.

\section{Background}

\subsection{FAIR and adjacent frameworks}

The FAIR principles \cite{wilkinson2016} have become the organising framework for research data stewardship. Findability, Accessibility, Interoperability, and Reusability map cleanly onto the technical and semantic layers of data management, and automated assessment tools, notably F-UJI \cite{devaraju2021}, enable evaluation of FAIR compliance at scale. In recent years, FAIR has become increasingly embedded in funder and publisher policies to ensure that publicly funded outputs are maximally accessed and used. Interpretation and implementation guidance have continued to evolve \cite{jacobsen2020}, and empirical FAIR assessments applied to the same datasets sometimes diverge \cite{krans2022}, both of which serve as reminders that FAIR is a living framework rather than a closed specification. Complementary frameworks extend their scope: the CARE (Collective benefit, Authority to control, Responsibility, Ethics) principles \cite{carroll2020} articulate collective benefit, authority to control, responsibility, and ethics for Indigenous data governance, and the TRUST (Transparency, Responsibility, User focus, Sustainability, Technology) principles \cite{lin2020} outline the requirements of trustworthy digital repositories.

FAIR is, by design, a stewardship framework and, as such, describes the dataset's properties and metadata. It does not describe how a specific user, with specific tools, skills, and context, experiences the process of finding, accessing, interpreting, and applying the data, nor the design choices providers make when balancing user convenience with service sustainability. This omission is a feature of FAIR's stated scope rather than a deficiency, yet it leaves a class of questions that operational providers encounter daily.

\subsection{Data friction in science and technology studies}

The term data friction has a prior history in science and technology studies on which the present framework builds. Edwards~\cite{edwards2010}, in a history of climate science and its data infrastructures, introduced the term to describe the energy consumed as data move between forms, formats, and disciplines; earlier work on infrastructures by Bowker and Star~\cite{bowker1999} had already shown how classification systems and metadata standards carry political and social consequences, and subsequent elaboration --- including Edwards et al.~\cite{edwards2011}'s account of science friction at disciplinary interfaces, and Borgman's situation of data within knowledge infrastructures \cite{borgman2015} --- extends the same lineage.

Bates~\cite{bates2018}, examining data friction in research data sharing and online communication, further develops the concept. Sites of data friction are political in Bates's account: they are shaped by the collective decisions of actors with unequal power, and friction may be something to enable and foster rather than to overcome. This insight is the direct intellectual antecedent of the engineered friction concept developed in Section 3.2. The contribution of the present paper is to operationalise that insight for operational data service providers: where Edwards and Bates describe and analyse friction, DFF offers providers a means to see it, score it, and act on it, including by deliberately retaining engineered friction where it serves sustainability, equity, or continued openness.

\subsection{Process-improvement and usability traditions}

Readers familiar with process-improvement methodologies will recognise a family resemblance between DFF and two established traditions: the Theory of Constraints \cite{goldratt1984} and Value Stream Mapping \cite{rother2003}. Both ask where, in an end-to-end system, effort accumulates. DFF inherits from the instinct to examine the entire journey rather than a single step. Two differences distinguish DFF. First, both traditions assume an orchestrator who owns the value stream end-to-end; a data reuser's journey is not orchestrated in this sense, since it crosses third-party tooling, community channels, and conventions set by external bodies. Second, both treat friction as something to reduce or remove; DFF distinguishes engineered from accidental friction, retaining the former where it serves sustainability or equity. DFF is also not a usability framework in the human-computer interaction sense \cite{nielsen1994,norman2013}: usability testing evaluates interface design, whereas DFF operates at the ecosystem level, encompassing documentation, format conventions, legal terms, community tooling, and support responsiveness.

\section{The Dataset Friction Framework}

The framework has three components: a set of six friction dimensions, a scoring approach for applying them, and an explicit distinction between engineered and accidental friction. It is built from a single underlying notion. We define a dataset friction factor as any attribute of a dataset, its delivery, or its surrounding ecosystem that increases the effort a user must expend to move from awareness of the dataset to productive use. Friction compounds across dimensions, in that low friction on any single dimension does not guarantee a low-friction experience overall; the question of how factors aggregate --- whether the user's experience is governed by the maximum friction on any one dimension (a bottleneck view) or by the sum across dimensions (a total-effort view) --- is intentionally left open. The framework presents dimension-level profiles rather than composite scores.

Two framing considerations apply. The first is that friction is not uniformly distributed across user types: a climate researcher with Python fluency encounters different friction than a local-government analyst, and a user operating a production service encounters different friction than one performing exploratory analysis. The framework therefore supports descriptions of friction for named user profiles rather than a single universal score --- in effect, the persona concept from interaction design, applied not to a single interface but to the whole data-service ecosystem. The second is that some friction is engineered, in the sense that it reflects deliberate provider choices in service of sustainability, equity, or reliability; the framework is designed to make both accidental and engineered friction visible.

\subsection{The six dimensions}

Table 1 presents the six dimensions with example factors, indicative low- and high-friction descriptors, and a scoring cue. The scoring cue identifies a diagnostic question that a reviewer with working knowledge of the dataset can apply.

\begin{longtable}[]{@{}
  >{\raggedright\arraybackslash}p{(\columnwidth - 8\tabcolsep) * \real{0.1922}}
  >{\raggedright\arraybackslash}p{(\columnwidth - 8\tabcolsep) * \real{0.2138}}
  >{\raggedright\arraybackslash}p{(\columnwidth - 8\tabcolsep) * \real{0.2137}}
  >{\raggedright\arraybackslash}p{(\columnwidth - 8\tabcolsep) * \real{0.2136}}
  >{\raggedright\arraybackslash}p{(\columnwidth - 8\tabcolsep) * \real{0.1668}}@{}}
\toprule\noalign{}
\begin{minipage}[b]{\linewidth}\raggedright
\textbf{Dimension}
\end{minipage} & \begin{minipage}[b]{\linewidth}\raggedright
\textbf{Example factors}
\end{minipage} & \begin{minipage}[b]{\linewidth}\raggedright
\textbf{Low-friction descriptor}
\end{minipage} & \begin{minipage}[b]{\linewidth}\raggedright
\textbf{High-friction descriptor}
\end{minipage} & \begin{minipage}[b]{\linewidth}\raggedright
\textbf{Scoring cue}
\end{minipage} \\
\midrule\noalign{}
\endhead
\bottomrule\noalign{}
\endlastfoot
1. Discoverability and Understanding & Documentation quality; catalogue integration; depth of onboarding path & Structured user-level guides; listed in open or federated catalogue; quickstart reachable in one or two clicks & Schema-only or absent documentation; not indexed in common catalogues; quickstart buried or absent & Time from landing page to a functioning example \\
2. Access and Delivery & Access method complexity; authentication overhead; delivery reliability and latency & Standards-based API (e.g. OGC EDR); anonymous or lightweight access; predictable performance & Bespoke protocol; institutional credentials required; variable performance & Number of steps to first successful data request \\
3. Licence and Legal & Licence clarity; attribution pathway; redistribution terms; eligibility constraints & Clear open licence (e.g. CC BY 4.0); single attribution string; no eligibility gate & Ambiguous or bespoke terms; manual attribution workflow; sector or geography restrictions & Whether a reuser can act on the licence terms without legal review \\
4. Data Structure and Format & Format openness; metadata completeness; temporal and spatial granularity fit & Open, widely tooled formats (NetCDF, Zarr, GRIB with ecCodes); CF-compliant metadata & Proprietary or sparsely documented formats; incomplete metadata; granularity that forces resampling & Whether a typical user can open, identify, and subset the data with standard tools \\
5. Tooling and Support & Client libraries; worked examples; support responsiveness; community presence & Maintained client libraries; notebooks and recipes; responsive support; active user community & No client libraries; no examples; self-serve only; no visible user community & Availability of a functioning example the user can adapt \\
6. Overall Complexity & Consistency across datasets; learning curve; transparency of updates & Consistent API, licence, and conventions across the catalogue; versioned, public change logs & Inconsistent approach across datasets; silent changes; training required before first use & Whether mastery of one dataset transfers to the next \\
\end{longtable}

\emph{\textbf{Table 1.} The Dataset Friction Framework: six dimensions with example factors, indicative descriptors, and scoring cues.}

\begin{figure}[htbp]
\centering
\includegraphics[width=\linewidth]{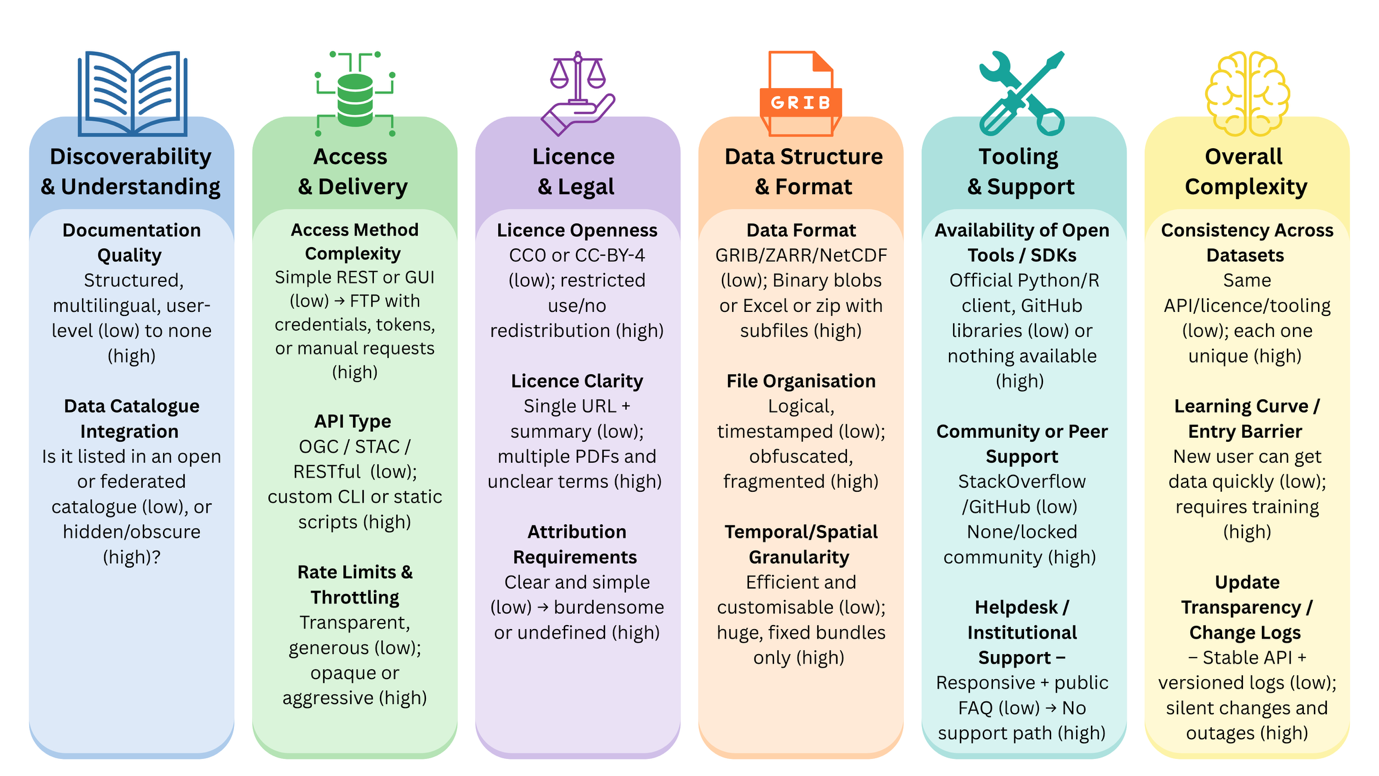}
\caption{The Dataset Friction Framework: six dimensions with example factors, indicative descriptors, and scoring cues (graphical summary).}
\label{fig:summary}
\end{figure}

\subsubsection{Dimension 1: Discoverability and Understanding}

A dataset must be found and sufficiently understood for a user to decide whether to invest further effort. Friction arises from weak or inconsistent documentation, limited catalogue integration, and onboarding paths that require repeated navigation to reach a working example. The diagnostic heuristic is the time it takes a new user to move from a landing page to a first piece of executable code. An example of high friction could be the need for a user to access 15 different documentation webpages to use a single piece of software to access a single dataset.

\subsubsection{Dimension 2: Access and Delivery}

Once a user has committed to using the data, friction depends on how the data are retrieved and on three sub-factors in particular: access method complexity, authentication overhead, and delivery reliability. Standards-based interfaces, such as the OGC Environmental Data Retrieval (EDR) API \cite{ogc2022}, reduce friction by allowing users to apply existing knowledge, whereas bespoke protocols increase it by requiring fresh learning for each provider.

\subsubsection{Dimension 3: Licence and Legal}

Licensing is addressed by the Reusability principle of FAIR, and yet the user-experience layer has its own characteristics. A clear, open licence such as CC BY 4.0 \cite{creativecommons2013}, paired with a single, copy-ready attribution string, is low-friction; bespoke terms, manual attribution workflows, and eligibility constraints that require institutional review increase friction even where the underlying intent is open. This dimension also surfaces the precondition layer of checks such as eligibility or security clearance that gate access before any other friction is encountered. However, connected with Dimension 1, if the licence conditions are not obvious or indicated alongside the dataset in question or held within the metadata itself, then user friction occurs in that the user has to go searching or ask the provider for the conditions of use -- a case still seen regularly within ECMWF and Copernicus support functions.

\subsubsection{Dimension 4: Data Structure and Format}

Format in isolation is a weak indicator of friction. A moderately complex format may be low friction if it is well tooled --- NetCDF, Zarr \cite{zarr2024}, and GRIB \cite{wmo2019} paired with ecCodes \cite{eccodes2024} are each widely tooled examples --- and a nominally simple format may be high friction if metadata are sparse or granularity does not fit common use. Tooling enters this dimension only as a property of the format's ecosystem --- whether mature libraries exist for it at all --- and is distinct from Dimension 5 (Tooling and Support), which concerns the client libraries, examples, and assistance the provider itself supplies. The operational question is whether a typical user can open, identify, and subset the data with standard tools, without recourse to a bespoke parser. However, friction in format is also apparent in the use of older formats, while newer user-friendly formats exist and are requested.

\subsubsection{Dimension 5: Tooling and Support}

Client libraries, worked examples, responsive support, and an active user community all reduce the effort required to use a dataset. The availability of a single functioning example that a user can adapt is often the decisive factor between a dataset that is widely used and one that is used only by its originators --- an observation with antecedents in the API-learnability literature \cite{robillard2011}.

\subsubsection{Dimension 6: Overall Complexity}

The final dimension captures friction visible only at the catalogue scale. Consistency across a provider's datasets --- the same API, licence, and conventions --- allows mastery of one dataset to transfer to the next, while inconsistency multiplies friction across a catalogue, and silent updates can undermine user trust even where individual datasets are well constructed. This dimension is particularly relevant in federated access environments, since as Europe's meteorological data spaces mature, users will increasingly encounter data from multiple providers presented through shared interfaces.

\subsection{Engineered friction: DFF as a design tool}

A framework that treats all friction as a defect misleads providers. Open data services are funded, staffed, and operated under real institutional, financial, and legal constraints, and three questions focus the design problem: what is the cost of zero friction, who pays to remove it, and when is friction necessary to ensure impact rather than merely access? Mature open data services typically carry friction that was deliberately introduced to sustain the service, to treat different user classes equitably, or to preserve the commercial channels that subsidise open provision. We term such friction engineered friction, building on Bates~\cite{bates2018}, who argues that friction may be something to enable and foster rather than to overcome.

Engineered friction is to be distinguished from accidental friction. Accidental friction includes the outdated documentation page, the undocumented API quirk, and the broken example notebook, and it is, by definition, a candidate for remediation. Engineered friction includes the embargoed release window, the registration gate, the coarsened resolution, and the differentiated support tier, and it is a provider choice --- we argue --- a defensible one, even where it is indistinguishable from accidental friction in a single user's experience. The contribution of the framework here is to give providers a vocabulary for naming deliberate choices and a structure for defending or revising them, rather than leaving those choices invisible.

A consistent menu of design axes recurs across open data services, in which each axis presents a lower-friction and a higher-friction option, and each carries a legitimate rationale for the higher-friction choice. Table 2 summarises nine such axes, drawn particularly from ECMWF's experience of operating both restrictive and open data channels \cite{pidduck2026}. Tiered access models of this kind are well established across public services --- healthcare co-pays, road tolls, university fee structures --- and the logic is the same: maintain universal access while enabling financial sustainability and prioritised services for specific user needs.

\begin{longtable}[]{@{}
  >{\raggedright\arraybackslash}p{(\columnwidth - 6\tabcolsep) * \real{0.2350}}
  >{\raggedright\arraybackslash}p{(\columnwidth - 6\tabcolsep) * \real{0.2671}}
  >{\raggedright\arraybackslash}p{(\columnwidth - 6\tabcolsep) * \real{0.2671}}
  >{\raggedright\arraybackslash}p{(\columnwidth - 6\tabcolsep) * \real{0.2308}}@{}}
\toprule\noalign{}
\begin{minipage}[b]{\linewidth}\raggedright
\textbf{Design axis}
\end{minipage} & \begin{minipage}[b]{\linewidth}\raggedright
\textbf{Lower-friction configuration}
\end{minipage} & \begin{minipage}[b]{\linewidth}\raggedright
\textbf{Engineered-friction configuration}
\end{minipage} & \begin{minipage}[b]{\linewidth}\raggedright
\textbf{Sustainability rationale}
\end{minipage} \\
\midrule\noalign{}
\endhead
\bottomrule\noalign{}
\endlastfoot
Latency of release & Real-time open release & Embargoed or delayed open release & Preserves a paid real-time channel that funds operations; also reduces system load and serving cost \\
Resolution and coverage & Full native resolution open & Coarsened, subset, or sampled open & Bounds storage and egress costs; retains a commercial product differentiated by resolution \\
Registration and identification & Anonymous access & Account-based access & Enables impact measurement, abuse mitigation, contact for material changes \\
Support tiering & Equal support for all users & Differentiated support by tier & Sustains operational SLA for paying users while preserving best-endeavours open support \\
Volume and rate limits & Unmetered access & Volume caps, rate limits, queueing & Manages infrastructure load; equalises access during peak demand \\
Format and tooling & Multiple ready-to-use formats & Single native format requiring client tooling & Bounds maintenance burden; relies on community tooling ecosystem \\
Attribution requirements & Optional attribution & Mandatory machine-readable attribution & Sustains impact recognition and licence compliance \\
Eligibility gating & Open to all & Sector or geography-restricted access & Honours upstream licence obligations or treaty terms \\
Versioning and deprecation & Continuous silent updates & Versioned releases with migration windows & Sustains downstream user reliability at the cost of release frequency \\
\end{longtable}

\emph{\textbf{Table 2.} A menu of recurring sustainability trade-offs in open data service design. Each axis presents a lower-friction and an engineered-friction configuration; the engineered configuration carries a legitimate sustainability rationale.}

A framework that cannot distinguish engineered from accidental friction will either treat all friction as waste --- and inadvertently dismantle the design choices through which open services are made durable --- or treat all friction as legitimate and forgive the accidental degradation of services that ought to be improved. Holding the two apart also changes the nature of the optimisation: the goal is not to minimise a single quantity --- friction treated as waste --- but to balance a more complex cost function in which some friction is deliberately retained where it sustains the service. The distinction is therefore foundational, as it is the analytic move that allows a DFF profile to be read as a diagnostic instrument rather than a leaderboard.

\section{Methods}

\subsection{Scoring approach}

For DFF to support comparison across datasets and providers, a scoring approach is required. The approach proposed here is deliberately lightweight. Its goal is for a reviewer with a working knowledge of a dataset to produce a DFF profile without a full technical audit, and for scores to be sufficiently anchored to support comparative use. Each dimension is assigned a score from 1 (low friction) to 5 (high friction). Profiles are presented as radar charts or dimension-level tables rather than being reduced to a single composite figure, since aggregating them into a single number obscures the dimension-level information providers most need.

To support consistent application and reduce rater variance, we developed a companion web-based scoring instrument implementing a Behaviourally Anchored Rating Scale (BARS; \cite{smith1963}) for each of the six dimensions. Behaviourally anchored scales replace vague qualitative descriptors with concrete behavioural statements tied to each scale point; they constrain rater interpretation more tightly than rubric tables alone and are the established method for improving inter-rater reliability in qualitative research instruments \cite{smith1963,woehr1994}. Each anchor describes an observable, checkable condition --- for example, the score-3 anchor for Access and Delivery reads ``free registration plus API token and/or asynchronous request-and-queue retrieval'' --- so that independent reviewers encounter the same decision criteria. Scoring is further anchored by the low-friction and high-friction descriptors in Table 1, supported by the scoring cues; where possible, the cue is a measurable proxy such as time to first successful request, number of steps from landing page to working example, or click depth to a quickstart, rather than a matter of reviewer judgement alone.

The instrument defines four user profiles: new-user, Python-literate scientist (the profile used in Table 3), GIS/portal analyst, and operational integrator. These make explicit the DFF principle that friction is not uniformly distributed across user types: the same dataset is experienced differently by a scripting researcher than by a portal analyst, a difference to surface rather than to average away. Each profile carries guidance on how the anchors should be read for that user --- for an operational integrator, delivery reliability and change transparency (D2, D6) dominate; for a GIS/portal analyst, Access and Tooling friction rise where scripted access is unavailable --- so that a reviewer scores the same dataset from a stated point of view rather than applying a single universal weighting. Profile-specific scoring is therefore a feature to report explicitly rather than a methodological limitation to suppress.

A scored profile may be annotated to distinguish engineered from accidental friction. The framework supports two complementary modes. The first is external provenance assessment, in which a reviewer classifies observed friction as engineered if the provider has published a rationale, and as accidental otherwise. The second is a provider-annotated assessment, in which the provider publishes a self-assessment, with each engineered score accompanied by a brief rationale. Accidental friction is flagged as a candidate for remediation; engineered friction is made explicit and defensible.

The scoring instrument is published as supplementary material to this paper. It is intended as a practical tool for data providers and dataset curators to evaluate their own services; feedback from its use in practice is invited and will inform subsequent calibration of the anchor descriptions.

\subsection{Empirical evidence base}

\textbf{Corpus construction.} We analysed all externally-visible user support tickets raised in two ECMWF Jira helpdesk projects between 1 January 2024 and 31 May 2026: the Data Services Customer project (DSC; n = 13,794), which handles licensed commercial users and operational data recipients, and the Copernicus Support at ECMWF project (CUS; n = 4,762), which handles open and registered users of the Copernicus Climate Data Store and Atmosphere Data Store. Together, these constitute a corpus of 18,556 tickets. A third project (SD, n $\approx$ 30,931 over the same period) handles internal infrastructure, security incidents, and hardware operations; SD tickets were excluded as they do not represent user-facing data friction events. Within the retained corpus, approximately 9\% of tickets (n $\approx$ 1,670) were identified as automated batch submissions created at exactly 00:05 UTC by a scheduled internal job; these are flagged in Section 5.2 but included in all dimension counts, which therefore represent conservative lower bounds on genuine user-initiated friction events.

\textbf{Label-based classification.} ECMWF support staff apply a label taxonomy to incoming tickets; where consistently applied, labels map directly to DFF dimensions: for example, portal-problem-downloading-data, slow-webmars, portal-problem-receiving-RT-data, portal-get-our-products, and portal-report-problem-on-computing map to Access and Delivery; and portal-username-password-or-authorised-access-issues, archive-commercial-renewal, portal-request-research-licence, and portal-commercial-licence map to Licence and Legal. However, labelling practice was demonstrably inconsistent over the study period: the label taxonomy evolved as new service types were introduced; different support staff applied labels in different ways; and manual inspection of sampled tickets revealed instances of systematic miscategorisation --- for example, technical tooling issues were assigned to licence-related labels. Approximately 57\% of DSC tickets and a substantial proportion of CUS tickets carry only a catch-all label (portal-other) or a dataset version identifier (e.g. ds-50r1) and are not captured by label queries at all. Label-based counts are therefore treated as independently derived conservative lower bounds per dimension, not as the primary evidence; free-text keyword search across ticket bodies is the primary classification method throughout.

\textbf{Free-text classification and query refinement.} Keywords were selected to target user-expressed friction language rather than incidental keyword occurrence. For each dimension, JQL queries were refined iteratively --- typically across four to six rounds --- with each candidate keyword set validated against a manually inspected random sample of at least 15 tickets drawn from the result set. Validated noise sources removed during refinement included: automated ECMWF PRC: Quote XXXXX system notifications (inflating Discoverability counts); commercial account-management tickets in the DSC project (inflating Data Structure and Format counts); and high-frequency generic terms (API, version) whose breadth made them unsuitable as dimension-specific signals. The final keyword sets are provided as Supplementary Table S1 to support replication. We deliberately did not apply unsupervised topic modelling approaches (e.g., Latent Dirichlet Allocation, Blei et al. 2003; BERTopic, Grootendorst 2022): these methods are appropriate for constructing a taxonomy from an unlabelled corpus, whereas our goal was to validate a pre-specified, theory-derived taxonomy against an independent corpus. Applying topic modelling to select which topics support dimensions derived from that same corpus would have been circular. Counts per dimension represent tickets containing at least one signal keyword; individual tickets frequently exhibit signals for multiple dimensions, so counts are not mutually exclusive and will sum to more than the corpus total. Reported figures should be read as indicative lower bounds --- tickets resolved silently or through community channels are not captured --- rather than as exhaustive measures of the prevalence of friction.

\textbf{Affect markers as a sentiment proxy.} Across all dimensions, tickets containing the terms ``frustrating'', ``unclear'', ``confused'', ``impossible'', ``difficult'', or ``annoying'' were counted as a simplified rule-based sentiment proxy (n = 1,205; 7\% of corpus), providing a conservative lower bound on negatively-valenced interactions. This approach is consistent with affect-marker detection methods applied to Jira and Stack Overflow corpora in empirical software engineering \cite{ortu2015,novielli2018}, though it does not produce a scored sentiment distribution. Application of a fully validated sentiment classifier --- such as VADER \cite{hutto2014}, developed for short social-media text and applied to software issue trackers --- would extend this analysis to a scored distribution and enable comparison across providers and time periods; this is noted as future work (Section 6.5).

\textbf{Additional evidence streams.} Three supplementary evidence streams were used to triangulate the ticket findings: (i) the ECMWF Open Data User Survey 2024 (n = 581 respondents), distributed to registered open data users and covering licence awareness, attribution practice, and discovery patterns; (ii) the ECMWF Licensed User Survey 2025 (n = 68 respondents), covering access methods, licence familiarity, and support experience among commercial licence holders; and (iii) analysis of open issues in two public GitHub repositories --- ecmwf/cdsapi (316 stars, 158 issues at time of analysis) and ecmwf/ecmwf-opendata (320 stars, 64+ issues) --- categorised against DFF dimensions following the issue classification approach of Antoniol et al.~\cite{antoniol2008} as adapted by Kallis et al.~\cite{kallis2019}. Issues were categorised by one author; a sample of 30 issues was independently reviewed by a second author. GitHub issue analysis provides evidence from a user population entirely independent of ECMWF's internal support infrastructure, confirming that observed friction patterns persist across channels and institutional boundaries.

\subsection{Toward automated DFF assessment}

The scoring approach described in Section 4.1 is reviewer-applied, and inter-reviewer variance is a known limitation. An automated implementation of DFF that is robust to this variance and applicable at the catalogue scale is therefore a natural next step, and the architectural choices it requires are themselves analytically informative. We sketch the proposed approach here; a working implementation and broader empirical validation are forthcoming in a companion paper.

The closest existing analogue is F-UJI \cite{devaraju2021}, the automated FAIR assessment tool. F-UJI is a passive harvester: it fetches a dataset's landing page and metadata, parses structured fields for compliance with the FAIR principles, and returns a compliance score. Passive harvesting is well suited to FAIR because FAIR is defined in terms of the structural properties of metadata that can be inspected without interacting with the data service. DFF, by contrast, measures user-facing friction --- a property that depends on what a user must do, not on what the metadata says. The natural operationalisation is therefore active simulation: the assessment instrument issues real HTTP requests, attempts anonymous downloads, traverses authentication flows, installs and exercises client libraries, and inspects licence text for redistribution clauses, recording at each step what was required of the user. Friction-type tagging --- engineered or accidental --- is then derived from probe semantics: an anonymous-access failure caused by a documented registration requirement is engineered; an anonymous-access failure caused by an undocumented server error is accidental.

The architectural distinction between passive harvesting and active simulation has a practical consequence beyond accuracy: passive harvesters are vulnerable to provider-side conditions that do not affect human users but do affect bots, while active simulators are not. Section 5.6 documents one such case empirically. A full implementation across all six dimensions, with calibration against the worked example in Section 5.4 and validation on a larger, more diverse dataset, is the subject of forthcoming work. Because active simulation generates automated traffic against third-party services --- including, as Section 5.6 shows, services that deliberately defend against it --- such an instrument must operate within published terms of service, rate-limit its requests to a level comparable to a single human user, and never harvest credentials or attempt to circumvent bot-mitigation controls; where a provider's policy precludes automated assessment, that condition is recorded as an engineered-friction signal rather than treated as an obstacle to overcome.

\subsection{Computational tools and reproducibility}

Jira ticket data were accessed programmatically via the Atlassian Jira REST API using a Model Context Protocol (MCP) server integration within the Claude Code agentic development environment \cite{anthropic2025}. MCP is an open protocol that enables language model assistants to interact with external data systems via structured, auditable tool calls. The Jira MCP server translates JQL queries into validated API requests and returns paginated results. All queries were issued as JQL COUNT or LIST requests against the live ECMWF Jira instance, with results reviewed and refined interactively. This approach enabled rapid iterative query development --- typically four to six rounds per dimension --- with immediate sample inspection to assess signal quality at each iteration. While the query syntax used here is specific to the Atlassian Jira platform, the keyword taxonomy --- provided in the Zenodo deposit (see Data Availability) --- is system-agnostic: other institutions can apply equivalent searches in any text-searchable helpdesk system using the same terms. The ECMWF project codes and internal label names are not required; the friction signals are in the language of the tickets, not in how they were routed. Counts will increase as new tickets are created after May 2026, but the methodology is fully transferable. The language-model integration assisted query authoring and sample inspection; the reported counts were produced by deterministic JQL COUNT queries executed against the live instance, not by model judgement.

The raw ticket data contain user contact details and cannot be published; however, the full set of final JQL queries, together with the iterative refinement history for each dimension, is provided as Supplementary Table S1. Investigators using other helpdesk systems can apply equivalent keyword searches using the taxonomy documented there, since the friction signals reside in the language of the tickets rather than in platform-specific query syntax. Where the host helpdesk system distinguishes between internal staff and external user reporters (whether by group membership, email domain, or organisation field), the comparator filter documented in Supplementary Table S1 Section~2.1 can be reproduced directly.

\emph{Reproducibility note on point-in-time figures.} Ticket counts reflect the corpus as constituted at 31 May 2026 and will grow as new tickets are created. F-UJI FAIR compliance scores for IFS/AIFS Open Data and ERA5 were obtained from the public F-UJI assessment interface (https://www.f-uji.net) in early 2026. GitHub repository statistics (star counts, issue counts) were recorded in May 2026. All figures are therefore point-in-time and should be interpreted accordingly.

\section{Results: Empirical Validation}

The framework is now set against operational evidence. We report friction signals across the six dimensions (Section~5.1) and their temporal distribution (Section~5.2), test whether they converge across independent evidence streams (Section~5.3), and develop a three-dataset worked example (Section~5.4) that grounds the comparison between FAIR compliance and DFF friction (Section~5.5) and the passive-assessment artefact it reveals (Section~5.6).

\subsection{Friction signals across the six dimensions}

Figure 2 summarises the ticket-level evidence per dimension. Throughout this section, label-confirmed counts denote tickets carrying a support label that maps to the dimension --- a conservative lower bound, as explained in Section~4.2 --- while free-text counts denote tickets whose body contains a dimension keyword. Discoverability friction was the largest single signal in the free-text corpus, followed by Licence and Legal, Access and Delivery, Tooling and Support, and Overall Complexity. Authentication and Performance sub-dimensions of Access and Delivery are shown separately. Two cross-cutting signals --- explicit failure language (``broken'', ``not working'') and affect markers (``frustrating'', ``unclear'') --- are shown as bars at the foot of the figure; these are not dimensions in themselves but mark the presence of accidental friction and qualitative intensity respectively.

To address whether the reported signals reflect genuine external user-facing friction or internal staff workflow, each dimension count was also evaluated against the subset of tickets raised by reporters who are not members of the ECMWF Jira staff group. Internal staff account for 12.3\% of the corpus overall; every dimension's internal-staff share falls below this baseline --- substantially so for the label-confirmed Access count (0.9\%) and the Data Structure and Format count (1.6\%) --- confirming that the dimensions are genuinely user-facing rather than artefacts of internal ticket flow. Full external-only counts are reported in Supplementary Table S1 and shown as the inner bars in Figure 2. It is notable, however, that user friction is not always an external factor; even within an organisation, different stakeholders can experience it in different ways.

\textbf{Table 1.} \emph{DFF-dimension friction signals: all reporters versus the external-user subset. Counts are live re-verifications (19 June 2026) against the corpus snapshot of 26 May 2026; the corpus total of record (18,556) is retained in the text. Every dimension's internal-staff share falls below the 12.3\% corpus baseline, confirming that the signals are user-facing. Bold rows are the six DFF dimensions, D2 sub-patterns, and the corpus total; bold figures in the final column mark the top three friction signals in the external-user corpus. Counts are not mutually exclusive.}

\begin{longtable}[]{@{}
  >{\raggedright\arraybackslash}p{(\columnwidth - 8\tabcolsep) * \real{0.2692}}
  >{\raggedright\arraybackslash}p{(\columnwidth - 8\tabcolsep) * \real{0.1731}}
  >{\raggedright\arraybackslash}p{(\columnwidth - 8\tabcolsep) * \real{0.1987}}
  >{\raggedright\arraybackslash}p{(\columnwidth - 8\tabcolsep) * \real{0.1603}}
  >{\raggedright\arraybackslash}p{(\columnwidth - 8\tabcolsep) * \real{0.1987}}@{}}
\toprule\noalign{}
\begin{minipage}[b]{\linewidth}\raggedright
\textbf{DFF dimension}
\end{minipage} & \begin{minipage}[b]{\linewidth}\raggedright
\textbf{All reporters}
\end{minipage} & \begin{minipage}[b]{\linewidth}\raggedright
\textbf{External users only}
\end{minipage} & \begin{minipage}[b]{\linewidth}\raggedright
\textbf{Internal share}
\end{minipage} & \begin{minipage}[b]{\linewidth}\raggedright
\textbf{\% of external corpus}
\end{minipage} \\
\midrule\noalign{}
\endhead
\bottomrule\noalign{}
\endlastfoot
D1 Discoverability and Understanding & 5,701 & 5,532 & 3.0\% & \textbf{34.0\%} \\
D2.0 Access and Delivery (labels) & 4,344 & 4,303 & 0.9\% & \textbf{26.5\%} \\
D2.a Performance sub-pattern & 787 & 736 & 6.5\% & 4.5\% \\
D2.b Authentication sub-pattern & 2,054 & 1,979 & 3.7\% & 12.2\% \\
D3 Licence and Legal & 5,512 & 5,173 & 6.2\% & \textbf{31.8\%} \\
D4 Data Structure and Format & 1,091 & 1,074 & 1.6\% & 6.6\% \\
D5 Tooling and Support & 3,262 & 2,943 & 9.8\% & \textbf{18.1\%} \\
D6 Overall Complexity & 1,561 & 1,497 & 4.1\% & 9.2\% \\
Accidental friction markers (cross-cutting) & 4,877 & 4,674 & 4.2\% & 28.8\% \\
Affect markers (cross-cutting) & 1,205 & 1,162 & 3.6\% & 7.1\% \\
\textbf{Corpus total} & \textbf{18,531} & \textbf{16,251} & \textbf{12.3\%} & \textbf{100.0\%} \\
\end{longtable}

\begin{figure}[htbp]
\centering
\includegraphics[width=\linewidth]{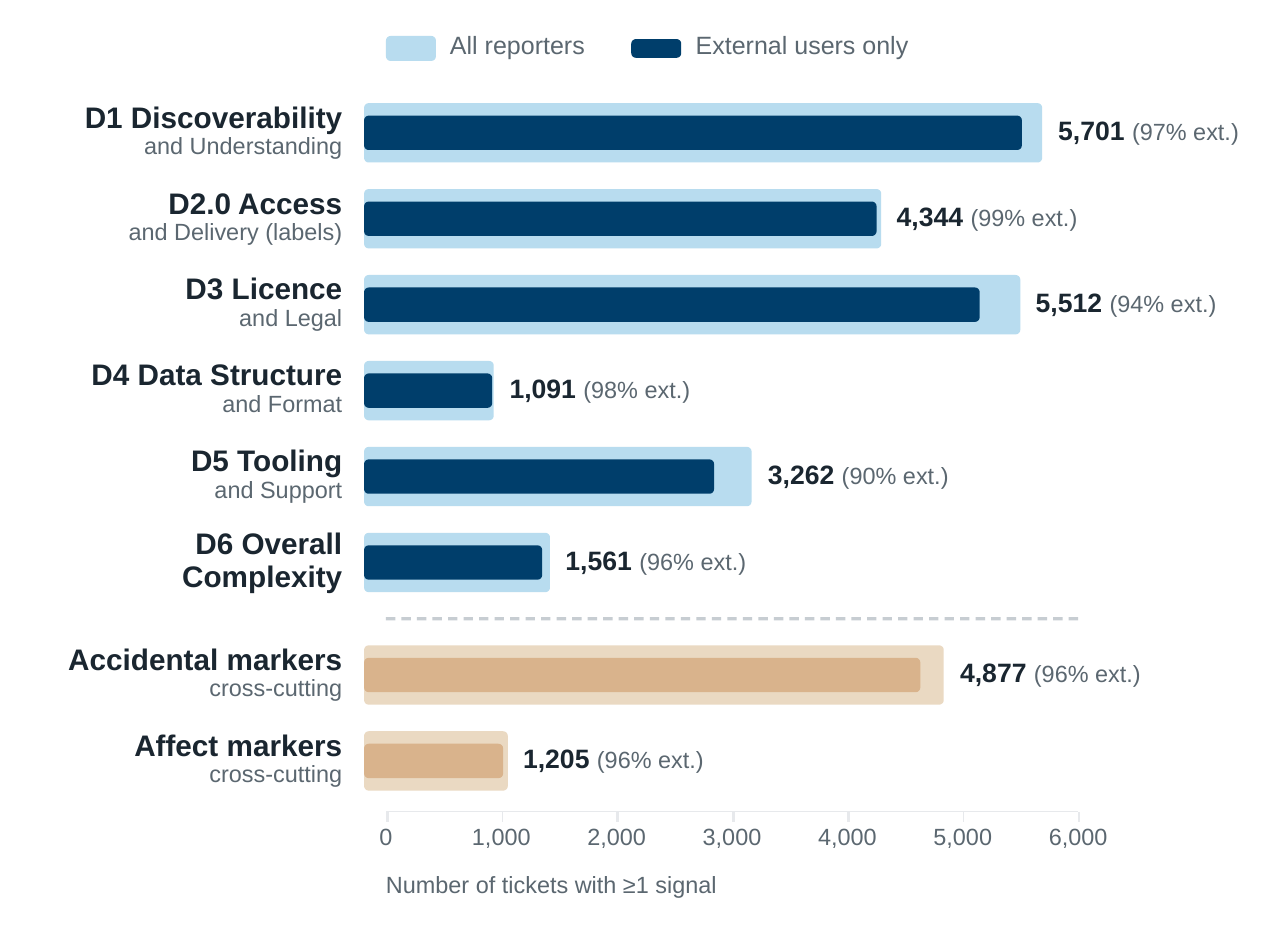}
\caption{DFF dimension friction signals from 18,556 user support tickets, ECMWF DSC and CUS projects, January 2024 to May 2026. Outer bars show all reporters; inner bars show the external-only subset (excluding the ECMWF Jira staff group; corpus internal share 12.3\%). Tickets frequently exhibit multi-dimensional friction, so counts are not mutually exclusive. Authentication and Performance are sub-dimensions of Access and Delivery. Cross-cutting signals (accidental friction markers and affect markers) appear below the dashed line and are not dimensions in themselves.}
\label{fig:bars}
\end{figure}

\subsubsection{Discoverability and Understanding}

Discoverability friction was the largest single signal in the free-text corpus, with 5,701 tickets (31\%) containing explicit navigational or existential queries --- users asking not how to process data, but whether it exists and where to find it. Representative examples include: ``Where can I find visibility data / how to compute visibility data'' (ECMWF user support ticket, July 2024); ``Where can I get the archived HRES and ENS data'' (April 2024); and ``Where can I download the 48-h delay S2S product of the ECMWF model?'' (March 2024). Of these, 97\% were raised by external users, confirming that the signal reflects user-facing friction rather than internal workflow. This pattern is consistent with survey findings: 43\% of open data users (ECMWF Open Data Survey 2024, n = 581) reported being unaware of the licence terms applicable to data they were actively using --- indicating that discoverability gaps extend from data assets to their associated conditions of use.

\subsubsection{Access and Delivery}

Label-confirmed access friction accounted for 4,722 tickets (25\%). A distinct performance sub-pattern appeared in 720 tickets containing explicit speed or latency language, including ``Era5-complete data downloading too slow'' (August 2024) and ``The download data speed is too slow, please help me with some solutions'' (August 2024). Authentication friction --- the largest identifiable sub-dimension of Access and Delivery --- accounted for a further 2,207 tickets (12\%), including multi-step API credential failures such as ``New CDS API Key does not authenticate'' (2024).

\subsubsection{Licence and Legal}

Licence friction appeared in 5,554 tickets (30\%), spanning attribution enquiries, commercial licence requests, and redistribution questions. A particularly clear instance of the interaction between engineered and accidental friction appeared in the ticket ``CDS API doesn't work despite accepting terms and license and updating'' (October 2024): the user had completed the mandatory licence-acceptance step --- an engineered friction gate --- but remained blocked by a downstream API failure, demonstrating that engineered friction mechanisms can amplify the impact of accidental events. At the survey level, 39\% of open data users (Open Data Survey 2024, n = 581) reported non-compliance with CC BY 4.0 attribution requirements --- confirming that Licence and Legal friction is not resolved by formal acceptance of terms.

\subsubsection{Data Structure and Format}

Data Structure and Format friction was the smallest signal in the ticket corpus, with explicit format-related queries appearing in approximately 1,143 tickets (6\%), representing the converged upper bound from five rounds of keyword query refinement; manual sampling of the result set suggests approximately half are clear genuine hits, with the remainder comprising incidental matches in operational DSC tickets. This dimension is under-represented in the support corpus relative to its likely population prevalence: format friction is typically resolved silently by users through documentation lookup or community channels rather than escalated to support, particularly where well-tooled formats (NetCDF, GRIB with ecCodes) carry an established ecosystem of community guidance. The survey evidence, where format is more frequently cited (see Section~5.3), is therefore the more reliable indicator for this dimension, and the ticket count should be read as a lower bound rather than an exhaustive measure.

\subsubsection{Tooling and Support}

Tooling friction --- identified through references to client APIs, Python libraries, notebooks, and documentation --- appeared in 3,262 tickets (18\%). Friction ranged from installation failures (``cdsapi installation'', August 2025) and version discontinuities (``Differences between files downloaded by version 0.6.1 and 0.7.0 of cdsapi'', August 2024) to documentation gaps (``Instructions are unclear on how to install and use CDS API on Windows'', October 2024). Qualitative intensity in this dimension occasionally reached acute levels; one ticket summarised the cumulative tooling experience as ``This is the most frustrating website I have ever dealt with'' (November 2024). This is not isolated: 1,205 tickets (6.5\%) across all dimensions contained explicit affect markers --- ``frustrating'', ``unclear'', ``confused'', ``impossible'', ``difficult'', ``annoying''.

\subsubsection{Overall Complexity}

The Overall Complexity dimension --- capturing friction from system-level inconsistency rather than any single access step --- appeared in 1,577 tickets (9\%), identified through references to breaking changes, deprecated interfaces, and unexpected behavioural differences between versions or endpoints. The cdsapi version discontinuity above illustrates how a single incident surfaces friction simultaneously at both the Tooling and Overall Complexity levels, consistent with the framework's recognition that dimensions are not mutually exclusive.

\subsubsection{Cross-cutting signal: accidental friction}

Across all dimensions, 4,877 tickets (26\%) contained explicit failure language (``broken'', ``not working'', ``doesn't work'', ``stopped working'', ``failed'', ``failure'', ``error''), providing a conservative lower bound on accidental friction events --- incidents that represent unintended defects rather than deliberate provider design choices. This distinction is directly operational within the DFF: accidental friction warrants remediation; engineered friction warrants a scored, transparent design decision.

\subsection{Temporal distribution of friction}

The temporal distribution of ticket creation provides independent evidence that best-endeavours support represents a genuine and quantifiable engineered friction trade-off (Figure 3). Of a stratified sample of 200 tickets, 9\% were automated batch submissions (created at exactly 00:05 UTC); of the remaining 182 genuine user-initiated tickets, 47\% were raised during European business hours (06:00--12:00 UTC), 27\% during the European afternoon and Americas morning (12:00--18:00 UTC), 13\% during the Americas evening (18:00--24:00 UTC), and 14\% between 00:00 and 06:00 UTC --- the Asia-Pacific working day --- when no synchronous support is available. The single most active hour was 08:00 UTC.

\begin{figure}[htbp]
\centering
\includegraphics[width=\linewidth]{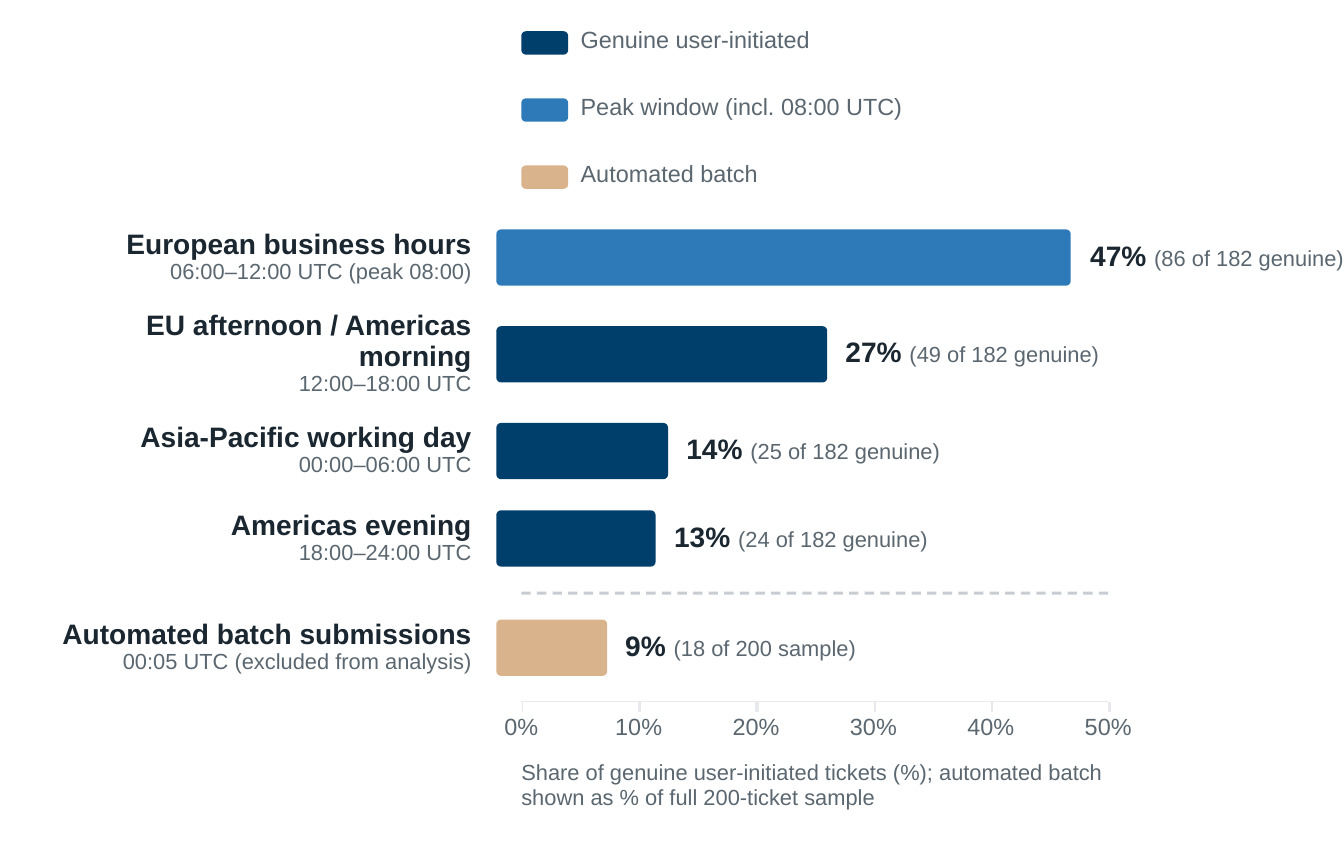}
\caption{Ticket creation time-of-day for a stratified sample of 200 tickets, in UTC. Of 200 tickets, 18 (9\%) were automated batch submissions at 00:05 UTC and are shown separately. Of the remaining 182 genuine user-initiated tickets, 47\% arrived during European business hours, 14\% during the Asia-Pacific working day. The 24-hour distribution confirms a globally dispersed user base for whom the current support model creates differential friction across geographies and time zones.}
\label{fig:temporal}
\end{figure}

The 24-hour distribution confirms a globally dispersed user base, for whom the current support model creates uneven friction across geographies and time zones. This is precisely the kind of trade-off the engineered/accidental distinction was designed to make visible: 24/7 synchronous support would be an operational commitment ECMWF does not currently fund, and the resulting friction for Asia-Pacific users is an engineered consequence of the chosen support model rather than an accidental defect.

\subsection{Convergence across independent evidence streams}

The friction patterns identified in the ticket corpus are independently corroborated across four additional evidence streams: two user surveys (Open Data Survey 2024, Licensed User Survey 2025), two public GitHub repositories (ecmwf/cdsapi, ecmwf/ecmwf-opendata), and the FAIR compliance assessment of the CDS infrastructure (Figure 3).

\begin{figure}[htbp]
\centering
\includegraphics[width=\linewidth]{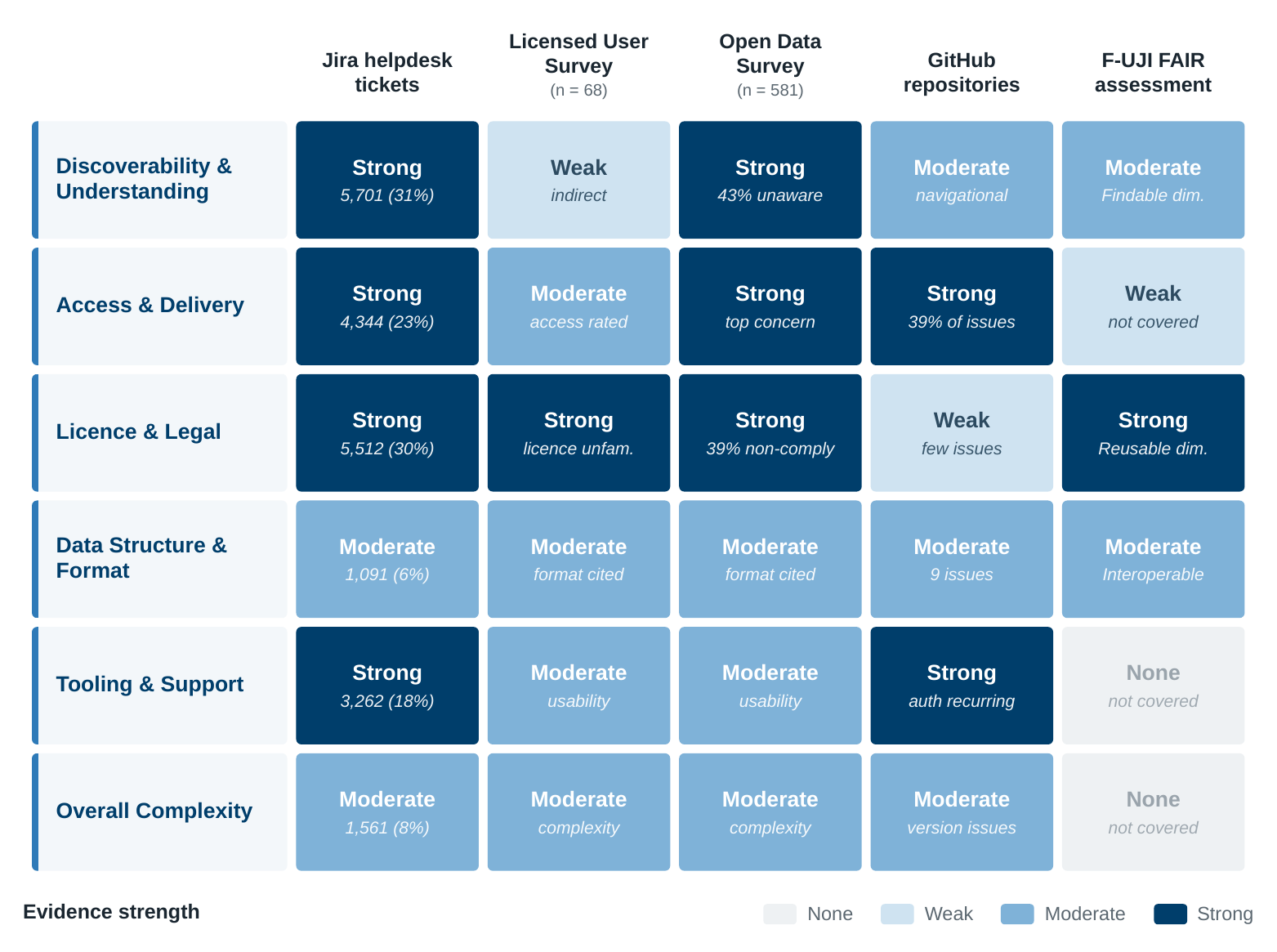}
\caption{Evidence convergence matrix. Evidence strength is shown across five independent sources for each of the six DFF dimensions: Jira helpdesk tickets, the ECMWF Licensed User Survey (n = 68), the ECMWF Open Data Survey (n = 581), public GitHub repositories (ecmwf/cdsapi, ecmwf/ecmwf-opendata), and the F-UJI FAIR compliance assessment. Five of the six dimensions are supported by at least three converging sources at moderate-to-strong strength.}
\label{fig:matrix}
\end{figure}

The ecmwf/cdsapi repository (316 stars; 158 issues since public release) shows 39\% of analysed issues addressing friction themes --- authentication failures and opaque error messages (10 issues), installation and environment setup (4 issues), and version discontinuities (4 issues). Critically, several issues are feature requests for fundamental usability improvements: ``Clearer errors in case of incorrect authentication'' (January 2025), ``Clearer error message when not agreed to licences'' (April 2025), and ``Add a `check authentication' function'' (August 2024). The recurrence of such requests, and their presence in a public repository entirely independent of ECMWF's support infrastructure, confirms that authentication and licence-gate friction persists across user populations, interaction channels, and institutional boundaries. The ecmwf/ecmwf-opendata repository (320 stars; 64+ issues) shows a parallel pattern, with format and access friction (9 issues) and version discontinuities (3 issues) prominent.

Figure 4 makes the convergence visible: five of the six DFF dimensions are supported by at least three converging evidence streams at moderate or strong strength. The remaining dimension (Data Structure and Format) shows moderate evidence across all streams, consistent with the explanation in Section~5.1 that this dimension appears less prominently in tickets than in user surveys.

\subsection{Worked example: three datasets, three FAIR--DFF relationships}

To make the framework concrete, we score three real datasets that occupy different points on the sustainability trade-off menu of Section 3.2 and exhibit three distinct FAIR--DFF relationships: the IFS/AIFS open data feed --- a real-time operational forecast product released under CC BY 4.0 via an anonymous HTTPS endpoint; ERA5 \cite{hersbach2020}, the ECMWF-produced reanalysis distributed through the Copernicus Climate Data Store (CDS; \cite{buontempo2022}) under CC BY 4.0 with user registration; and a representative German weather and climate dataset (G\_D5M, daily climate observations from the German national meteorological service's open geoportal). All three are open in the licence sense, and the IFS/AIFS feed and ERA5 both score 92\% against the F-UJI automated FAIR assessment \cite{devaraju2021}. Their DFF profiles, however, differ in ways that FAIR does not surface and that the sustainability menu anticipates. Scores below reflect the authors' expert judgement on publicly observable features of each service as of early 2026, applied under a `new-user, Python-literate scientist' profile.

\begin{longtable}[]{@{}
  >{\raggedright\arraybackslash}p{(\columnwidth - 6\tabcolsep) * \real{0.3205}}
  >{\raggedright\arraybackslash}p{(\columnwidth - 6\tabcolsep) * \real{0.2265}}
  >{\raggedright\arraybackslash}p{(\columnwidth - 6\tabcolsep) * \real{0.2265}}
  >{\raggedright\arraybackslash}p{(\columnwidth - 6\tabcolsep) * \real{0.2265}}@{}}
\toprule\noalign{}
\begin{minipage}[b]{\linewidth}\raggedright
\textbf{Dimension}
\end{minipage} & \begin{minipage}[b]{\linewidth}\raggedright
\textbf{IFS/AIFS Open Data}
\end{minipage} & \begin{minipage}[b]{\linewidth}\raggedright
\textbf{ERA5 via CDS}
\end{minipage} & \begin{minipage}[b]{\linewidth}\raggedright
\textbf{DWD G\_D5M}
\end{minipage} \\
\midrule\noalign{}
\endhead
\bottomrule\noalign{}
\endlastfoot
1. Discoverability & 1 & 2 & 2 \\
2. Access and Delivery & 1 & 3 & 1 \\
3. Licence and Legal & 1 & 1 & 1 \\
4. Data Structure and Format & 2 & 2 & 2 \\
5. Tooling and Support & 1 & 2 & 3 \\
6. Overall Complexity & 2 & 3 & 2 \\
\end{longtable}

\emph{\textbf{Table 3.} Illustrative DFF scores for the IFS/AIFS open data feed, ERA5 via the Copernicus Climate Data Store, and the DWD G\_D5M open geoportal dataset. Scores on a 1 (low friction) to 5 (high friction) scale under a new-user, Python-literate scientist profile. ERA5's Licence score of 1 reflects structural licence clarity (CC BY 4.0, single attribution string, no eligibility gate); user awareness of licence terms is a Discoverability (D1) signal rather than a Licence (D3) signal, and the survey evidence of low awareness is captured under Discoverability in Section~5.1.}

\begin{figure}[htbp]
\centering
\includegraphics[width=\linewidth]{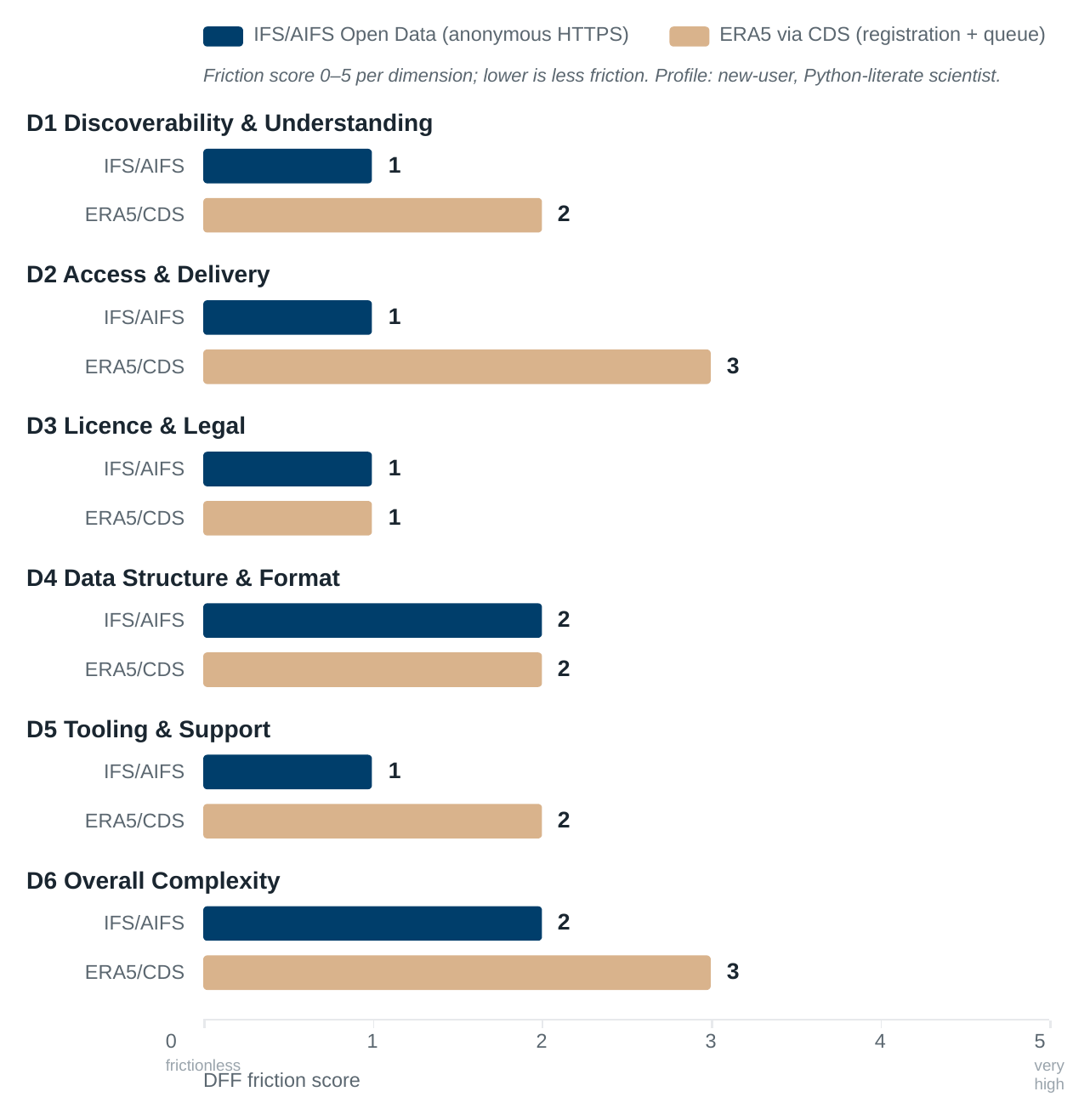}
\caption{Illustrative DFF profiles of the IFS/AIFS open data feed (solid) and ERA5 via the Copernicus Climate Data Store (dashed). The two profiles are closest on Licence and Legal and Data Structure, and diverge most on Access and Delivery and Overall Complexity, reflecting the engineered friction of CDS registration and queueing versus the anonymous HTTPS access model of the IFS/AIFS open data feed. A third dataset (DWD G\_D5M) is discussed in Section 5.6 and will be added to this profile comparison in the companion implementation paper.}
\label{fig:profiles}
\end{figure}

Three observations follow from the case. First, the framework discriminates between the services on the dimensions where they genuinely differ --- the IFS/AIFS feed and ERA5 diverge on Access and Delivery and Overall Complexity while agreeing on licence and format; DWD G\_D5M agrees with the IFS/AIFS feed on Access (anonymous HTTPS) but diverges on Tooling and Support, reflecting the absence of a maintained Python client library for the geoportal. Second, the engineered--accidental distinction can be applied concretely: the ERA5 registration step is engineered and defensible on the sustainability-menu grounds of Section 3.2, CDS request latency is accidental and is a candidate for remediation, and the DWD tooling gap is best understood as a gap to be filled rather than a design choice. Third, DFF is not a leaderboard: the three services serve different operational tasks, and the framework's output is a set of profiles that show where each service sits relative to the design menu --- the information providers need to review their own choices.

\subsection{FAIR compliance and DFF as non-redundant measures}

Both ECMWF datasets in the worked example achieve an F-UJI FAIR compliance score of 92\% \cite{ecmwf2026a}, yet they exhibit substantially different DFF profiles. IFS/AIFS Open Data carries a DFF Access score of 1 (anonymous HTTPS access, no registration); ERA5 via the CDS carries a score of 3 (registration, API token, request queueing). The same FAIR score, the same provider, but very different user-facing friction on the dimension that should most directly correspond to access. This disjunction is central to DFF\textquotesingle s contribution: FAIR compliance is a necessary but not sufficient condition for friction-free data access. FAIR addresses the structural properties of data and metadata; DFF addresses the user-facing cost of accessing them. The two measures are non-redundant, and the non-redundancy is not an abstract claim --- it is visible in real datasets from a single provider.

\begin{figure}[htbp]
\centering
\includegraphics[width=\linewidth]{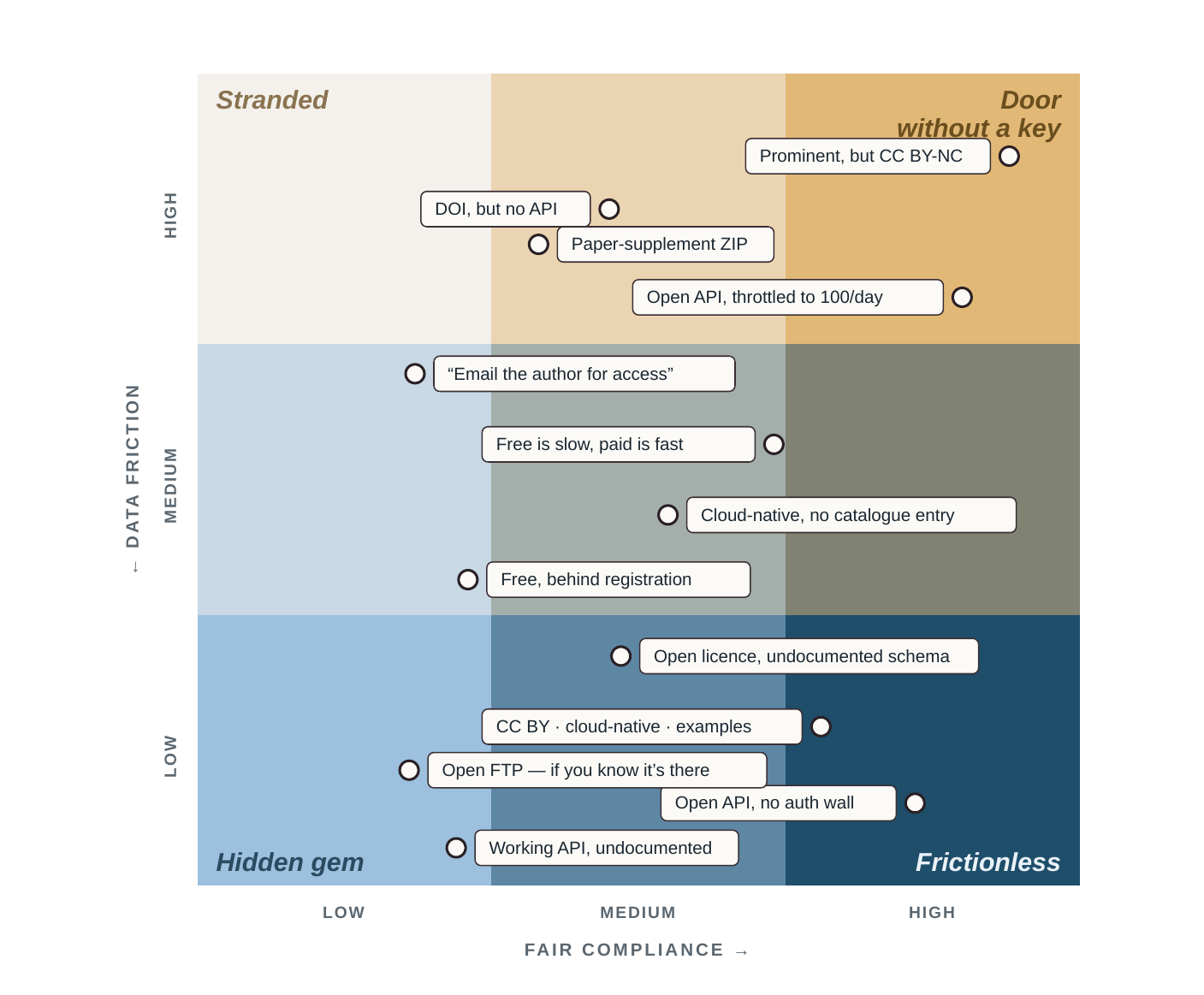}
\caption{The open-data landscape as a coordinate space rather than a binary. Datasets are positioned by FAIR compliance (horizontal) and user-facing data friction (vertical), the two properties that jointly determine whether an openly licensed dataset is actually reused. The two measures are non-redundant: a dataset can score well on one and poorly on the other. The four corners name characteristic modes. A \emph{frictionless} dataset combines high FAIR compliance with low friction; a \emph{hidden gem} is low-friction but poorly discoverable; a dataset behind a \emph{door without a key} is highly FAIR-compliant yet high-friction; and \emph{stranded} datasets are weak on both. The plotted positions are illustrative archetypes rather than measured scores, and most real datasets sit far from the frictionless corner that their FAIR compliance alone might imply.}
\label{fig:bivariate}
\end{figure}

\subsection{The DWD case: passive FAIR assessment and measurement artefacts}

The third dataset in the worked example reveals a further dimension of the FAIR--DFF relationship. It is, by the DFF account, a strong example of a low-friction open dataset: a human user reaches the data quickly and without barriers. The discrepancy examined here lies entirely in how an automated FAIR-assessment tool reads the service, not in the quality of the data or its stewardship. DWD G\_D5M, an open daily climate observation dataset distributed through the German national meteorological service's geoportal, returns a F-UJI FAIR compliance score of 42\% --- a score that would conventionally indicate poor metadata stewardship and incomplete openness. Manual inspection of the dataset by the authors, applying the DFF scoring approach of Section 4.1, returns a low-friction profile across most dimensions: anonymous HTTPS access succeeds, the licence is openly available and machine-readable, and the data are retrievable in a documented format. A human user encounters no access barrier of the kind that the 42\% FAIR score would suggest.

The cause of the discrepancy is not the dataset but the interaction between a passive harvester and the provider's infrastructure, and F-UJI reports it directly. Its assessment log for the geoportal records repeated HTTP 403 responses denying the harvester's user agent --- ``the host denied the User-Agent used (web scraping detection), retrying'' --- warns that ``connection problems have occurred that may influence the assessment result'', and notes that the resource ``does not identify itself as a `dataset' but as {[}`website'{]}, so F-UJI may not be the right tool for this type of resource.'' The 42\% therefore reflects the provider's bot-mitigation policy and a tool--resource mismatch, not the stewardship of the data, which a human user retrieves without difficulty. The full assessment record and diagnostic screen are provided as supplementary material. To the authors' knowledge, following a search of the F-UJI documentation and the published FAIR-assessment literature, this is the first documented instance of provider-side bot-mitigation producing this measurement artefact. The structural argument --- that any passive harvester is vulnerable to provider-side conditions that distinguish automated clients from human users --- follows from the architecture of passive harvesting and does not depend on the case being unique, and the resulting compliance scores can systematically misrepresent datasets that are well stewarded but defended against automated traffic.

Three observations follow. First, the FAIR--DFF relationship is non-monotonic: a high FAIR score can coexist with substantial friction (the ERA5 case), and a low FAIR score can coexist with low friction (the DWD case). FAIR compliance, as currently measured by leading passive harvesters, is neither a sufficient nor a reliable indicator of user-facing friction. Second, the DWD case strengthens the methodological argument for active-simulation assessment introduced in Section 4.3: an instrument that interacts with the service as a human user does is robust to anti-scraping policies in a way that a passive harvester is not and would not have produced the 42\% artefact. Third, the case is itself an instance of engineered friction --- bot-mitigation is a deliberate design choice with a legitimate rationale --- but it is engineered against a different actor (automated harvesters) than the friction the framework typically describes. The friction is real, but it targets the assessment instrument rather than the user, and the consequence is a measurement artefact rather than a usability barrier.

\begin{tcolorbox}[colback=boxfill,colframe=boxrule,boxrule=1pt,arc=2pt,left=10pt,right=10pt,top=8pt,bottom=8pt]
\textbf{Aligning FAIR and DFF: actions for the three worked-example datasets}\par\medskip
FAIR compliance and user-facing friction need not move together. For each dataset, alignment of the two measures implies a specific, deliberate choice rather than a universal fix.\par\medskip
\begin{itemize}[leftmargin=1.2em,itemsep=4pt,topsep=2pt]
\item \textbf{IFS/AIFS Open Data (FAIR 92\%, friction 1): already aligned.} Anonymous access delivers both a high FAIR score and low friction. The action is to hold the anonymous model and to record, on the sustainability menu of Section 3.2, the operational cost that this openness carries, so that the choice remains a conscious one as demand grows.
\item \textbf{ERA5 via the CDS (FAIR 92\%, friction 3): misaligned through access cost.} The same FAIR score coincides with registration, an API token, and request queueing. Two routes restore consistency. Removing registration would lower friction toward the IFS profile but raises the cost and load that registration is there to manage. Alternatively, the registration and queueing can be documented as engineered friction with a stated rationale, so that the friction reads as a transparent sustainability choice rather than an unexplained barrier. Request latency, by contrast, is accidental and is a candidate for remediation.
\item \textbf{DWD geoportal (FAIR 42\% as measured, low friction): the dataset needs no change.} Here the misalignment is in the measurement, not the data. The action sits with the assessment ecosystem: an active-simulation assessment that reaches the service as a human user does, together with a declared resource type, would resolve the artefact. The provider may additionally choose to expose dataset-typed metadata for harvesters, but this is optional and does not affect the low-friction user experience.
\end{itemize}\medskip
\textbf{Across a single provider, consistency carries its own implication:} for ECMWF's IFS and ERA5 to report the same friction as well as the same FAIR score, both would need the same access model. Either both move to anonymous access (maximising openness and minimising friction, at higher and less predictable cost), or both adopt registration-based access (accepting a small, transparent increment of engineered friction in exchange for a sustainable, manageable service). The framework does not prescribe which; it makes the trade-off explicit so that the choice can be made deliberately.
\end{tcolorbox}

\section{Discussion}

\subsection{DFF as a theory of change}

The framework implies a theory of change. Applied to an existing service, DFF makes friction visible by naming and scoring it across six dimensions, and the engineered--accidental distinction then partitions that friction into two actionable categories: accidental friction is identified as a candidate for targeted remediation, while engineered friction is surfaced, documented, and defended against the sustainability rationale that introduced it. Over iterative cycles of assessment, the gap between a dataset being published and a dataset being usable narrows.

The feedback loop is the key mechanism. Providers who score their own services and who invite external scoring gain a structured basis for prioritising improvements, and funders who require DFF profiles alongside FAIR assessments gain a more complete view of what has been built and of what it costs users to use it. This is not a claim that DFF adoption will, by itself, reduce friction across the ecosystem; it is a claim that visibility is a precondition for improvement, and that naming friction --- particularly the friction providers have chosen to retain --- is the first step toward a more deliberate and defensible open data landscape.

\subsection{The non-monotonic FAIR--DFF relationship}

The three-dataset worked example demonstrates that the relationship between FAIR compliance and DFF friction is non-monotonic. ERA5 shows the expected mode of disagreement --- high FAIR, high friction --- in which structural compliance does not predict user-facing cost. DWD shows a less anticipated mode --- low FAIR, low friction --- in which the low FAIR score is an artefact of how passive assessment instruments interact with provider-side bot policy, rather than a reflection of the dataset itself. The methodological consequence is that FAIR compliance scores, particularly when produced by passive harvesters, cannot be treated as a proxy for usability in either direction, and that an active-simulation approach (Section 4.3) is necessary both to measure friction accurately and to detect the class of measurement artefact illustrated by the DWD case. The empirical demonstration of this artefact is, to the authors' knowledge, the first of its kind in the literature on automated FAIR assessment.

\subsection{Audiences and applications}

DFF is intended for three audiences. For data providers, the framework offers a structured means of identifying accidental friction that is a candidate for remediation and engineered friction introduced for good reasons, and of revisiting whether the latter continues to earn its place in the service. For data reusers, DFF provides a vocabulary for describing what is difficult about a dataset and a basis for selecting between candidates. For funders and policymakers, the framework provides a lens on the gap between FAIR compliance and actual usability, and a basis for setting expectations of open data services beyond the FAIR baseline.

\subsection{Limitations}

The framework has acknowledged limitations. The first concerns reviewer subjectivity and inter-rater reliability: the scoring approach in Section 4.1 is reviewer-applied. The behaviourally anchored rubric in the companion scoring instrument is designed to constrain rater variance by providing concrete behavioural criteria at each scale point \cite{smith1963,woehr1994} but does not eliminate it. Formal inter-rater reliability statistics (Cohen's $\kappa$; \cite{cohen1960,landis1977}) have not been computed for the present paper; the instrument was calibrated to reproduce the reference scores in Table 3. This calibration establishes internal consistency between the instrument and the worked example; independent validation --- in which reviewers score datasets without access to the authors' scores --- is required to establish construct validity, and a prospective study reporting Cohen's $\kappa$ per dimension across independent reviewers is identified as a priority for future work.

The second concerns the classification methodology. The free-text keyword classification of support tickets was iteratively refined but not validated against a fully independent coding scheme. Established automated approaches --- including BERT-based issue-type classifiers \cite{kallis2019} and unsupervised topic models \cite{blei2003,grootendorst2022} --- were not applied; as noted in Section 4.2, this was a deliberate methodological choice because DFF dimensions are theory-derived rather than data-discovered, and applying topic modelling for validation would have been circular. Automated classification of a held-out ticket sample against the DFF dimension set is noted as future work and would yield formal precision and recall estimates for the keyword approach reported here.

The third is that the framework does not at present include a dedicated precondition meta-dimension; eligibility and security gating are surfaced within Licence and Legal, which is sufficient for current use cases, and a dedicated precondition layer is a candidate extension, particularly for services operating under mixed open and restricted-access regimes where the precondition check is the dominant source of friction.

The fourth is that the worked example covers three datasets across two providers. The three were selected to illustrate three distinct FAIR--DFF relationships rather than to constitute a statistical sample; cross-provider validation at scale, on a larger and more diverse dataset sample, requires the automated implementation and is part of forthcoming work.

The fifth is that the DWD anti-scraping artefact reported in Section 5.6 is, at present, a single documented case. The structural argument it supports --- that any passive harvester is vulnerable to provider-side bot policy --- follows from the architecture of passive harvesting rather than from the single case, but systematic measurement of the prevalence of this artefact across providers requires the automated implementation and is similarly part of forthcoming work.

The sixth is that provider-annotated scoring carries an obvious bias: providers scoring their own services have an incentive to classify friction as engineered rather than accidental, and to under-score accidental friction generally. This limitation is mitigated, though not eliminated, by the external-provenance mode.

The seventh concerns provider positionality: the institution that provides the validation corpus is also the one whose design choices are analysed as engineered friction. The authors have an interest in a framework that classifies deliberate provider choices as legitimate rather than defects; this is acknowledged, and application of DFF by independent assessors to ECMWF's services --- and by ECMWF staff to other providers' services --- is noted as a check on this bias.

The eighth is that the six dimensions are grounded in operational practice and in the science and technology studies literature cited in Section 2.2, rather than in a formal derivation, and a sensitivity analysis of the dimension set is future work.

\subsection{Future directions}

Four extensions of this work are planned or underway. First, the framework will be complemented by a companion paper introducing an automated assessment instrument that applies DFF at the catalogue scale by simulating the user journey directly rather than harvesting metadata (sketched in Section 4.3); that work will report implementation across all six dimensions, calibration against the Table 3 reference scores, and cross-provider application to a larger and more diverse dataset sample. Second, the companion scoring instrument published with this paper is being developed into a publicly hosted community platform at sites.ecmwf.int, through which providers, curators, and reusers can score services against the six dimensions and contribute their assessments to an openly aggregated corpus; beyond supporting community use and feedback, this platform is the mechanism by which the inter-rater reliability study identified in Section 6.4 becomes tractable, since accumulated independent assessments of the same datasets supply the multi-rater data required to compute Cohen's $\kappa$ per dimension \cite{cohen1960,landis1977} and to recalibrate the behavioural anchors against observed rating dispersion. Third, the ticket-corpus analysis will be extended longitudinally, tracking dimension-level counts against dated service-change events --- API migrations, documentation overhauls, client-library releases --- to provide a direct friction-reduction signal that lets providers measure the impact of targeted interventions. Fourth, the affect-marker proxy of Section 4.2 will be supplemented by a validated sentiment classifier \cite{hutto2014,ortu2015} to yield a scored distribution of affective intensity across dimensions and over time.

\section{Conclusion}

FAIR has given the open data community a shared language for stewardship; DFF is proposed as a complementary language for user experience and for service design. The two are mutually reinforcing: a dataset that scores well on both is genuinely useful and genuinely open, and a provider that applies both has a clearer view of what has been built and what it costs users to use it.

The reframing of DFF as a design tool, rather than an evaluation lens alone, is the contribution most directly relevant to providers. Mature open data services involve choices about immediacy, resolution, registration, and support tiering, and these choices are not defects to be engineered away toward a frictionless ideal but are the mechanisms through which open provision is made durable; DFF offers providers a language for these choices and a structure within which to review whether each remains appropriately calibrated. The empirical evidence presented in Section 5 --- 18,556 support tickets, two user surveys, two public software repositories, and three real datasets exhibiting three distinct FAIR--DFF relationships --- demonstrates that user-facing friction is a measurable and persistent feature of even a high-performing, FAIR-compliant open data infrastructure, and that the relationship between FAIR compliance and friction is non-monotonic in directions that current passive assessment instruments do not detect.

The validation presented here is necessarily situated: 18,556 tickets from one provider cannot establish DFF as a universal instrument but establish it as a useful one at a scale of evidence not typical for framework introduction papers --- the original FAIR principles paper \cite{wilkinson2016} was published without empirical validation data. Cross-institutional application --- by other providers, using the scoring instrument published here --- is the step that will determine how well the dimensions and anchors generalise across institutional contexts.

The methodological implication --- that active simulation of the user journey is required to assess dataset friction reliably, and to detect the class of measurement artefact that arises when passive harvesters meet provider-side bot policy --- is the subject of a companion paper currently in preparation, which will report the implementation, calibration, and cross-provider application of an automated DFF assessment instrument. The objective of this line of work, taken together, is modest but important: to make the gap between a dataset being published and a dataset being usable visible, comparable, and improvable, while giving providers credit for the deliberate design choices that sustain openness.

\section*{Data Availability}

The raw support-ticket corpus contains user contact details and is not publicly available in raw form. The keyword taxonomy used for free-text classification, the aggregated dimension-level ticket counts (with and without the external-user comparator filter), the iterative refinement history for each DFF dimension, and the survey instruments and aggregated results for the ECMWF Open Data Survey 2024 (n = 581) and Licensed User Survey 2025 (n = 68) are deposited together at \url{https://doi.org/10.5281/zenodo.20759544}

No personally identifiable information is contained in the deposit; raw ticket bodies and reporter identities remain confidential. Public repository data (ecmwf/cdsapi, ecmwf/ecmwf-opendata) are openly accessible at https://github.com/ecmwf/cdsapi and https://github.com/ecmwf/ecmwf-opendata.

\section*{Code Availability}

The DFF Anchored Scoring Tool (dff\_scoring\_tool.html), a self-contained web-based instrument implementing the behaviourally anchored rating scale for all six DFF dimensions, is provided as Supplementary Software S1. No other custom code was produced during this study. Free-text classification used standard JQL keyword matching; the full set of queries is provided in Supplementary Table S1.

\section{Supplementary Table S1 --- DFF keyword taxonomy, JQL queries, and refinement history}

This supplement is the deposit-ready record of the keyword taxonomy, JQL queries, and refinement history used to classify the support-ticket corpus described in Section~4.2 and Section~5.1. Each dimension's final keyword set is given as a verbatim JQL fragment so that an independent investigator with Jira access to a comparable corpus can reproduce the dimension-level counts. The taxonomy is system-agnostic: the underlying terms can be applied as equivalent text searches in any text-indexed helpdesk system. Counts were live verified on 19 June 2026 against the corpus snapshot of 26 May 2026.

\subsection{S1.1 Corpus definition and external-user comparator}

\textbf{Corpus:} project in (DSC, CUS) AND created \textgreater= "2024-01-01" AND created \textless= "2026-05-26". Internal ECMWF staff reporters are members of the Jira group ecmwf; the external-only count for each dimension applies the additional filter reporter not in membersOf("ecmwf"). Corpus baseline: n\_total = 18,531; n\_external = 16,251; internal-staff share = 12.3\%. Every dimension's internal-staff share falls below this baseline, confirming that the reported signals are genuinely user-facing.

\subsection{S1.2 Master summary}

\textbf{Table S1.1.} \emph{Final keyword sets, live-verified counts (all reporters and external-only), internal-staff share, paper-of-record counts, and drift. The corpus total of record (18,556) is retained in the manuscript text; the live re-verification total (18,531) reflects 24 days of archival drift.}

\begin{longtable}[]{@{}
  >{\raggedright\arraybackslash}p{(\columnwidth - 12\tabcolsep) * \real{0.2457}}
  >{\raggedright\arraybackslash}p{(\columnwidth - 12\tabcolsep) * \real{0.1261}}
  >{\raggedright\arraybackslash}p{(\columnwidth - 12\tabcolsep) * \real{0.1261}}
  >{\raggedright\arraybackslash}p{(\columnwidth - 12\tabcolsep) * \real{0.1154}}
  >{\raggedright\arraybackslash}p{(\columnwidth - 12\tabcolsep) * \real{0.1154}}
  >{\raggedright\arraybackslash}p{(\columnwidth - 12\tabcolsep) * \real{0.1346}}
  >{\raggedright\arraybackslash}p{(\columnwidth - 12\tabcolsep) * \real{0.1368}}@{}}
\toprule\noalign{}
\begin{minipage}[b]{\linewidth}\raggedright
\textbf{Dimension / signal}
\end{minipage} & \begin{minipage}[b]{\linewidth}\raggedright
\textbf{n\_total}
\end{minipage} & \begin{minipage}[b]{\linewidth}\raggedright
\textbf{n\_external}
\end{minipage} & \begin{minipage}[b]{\linewidth}\raggedright
\textbf{Internal \%}
\end{minipage} & \begin{minipage}[b]{\linewidth}\raggedright
\textbf{Paper n}
\end{minipage} & \begin{minipage}[b]{\linewidth}\raggedright
\textbf{Drift}
\end{minipage} & \begin{minipage}[b]{\linewidth}\raggedright
\textbf{Sample / note}
\end{minipage} \\
\midrule\noalign{}
\endhead
\bottomrule\noalign{}
\endlastfoot
\textbf{D1 Discoverability} & 5,701 & 5,532 & 3.0\% & 6,888 & -17.2\% & PRC-Quote noise now excluded \\
\textbf{D2.0 Access (labels)} & 4,344 & 4,303 & 0.9\% & 4,722 & -8.0\% & label retirement + archival \\
D2.a Performance & 787 & 736 & 6.5\% & 720 & +9.3\% & speed / latency sub-pattern \\
D2.b Authentication & 2,054 & 1,979 & 3.7\% & 2,207 & -6.9\% & credential sub-pattern \\
\textbf{D3 Licence and Legal} & 5,512 & 5,173 & 6.2\% & 5,554 & -0.8\% & best-reproduced dimension \\
\textbf{D4 Data Structure \& Format} & 1,091 & 1,074 & 1.6\% & 1,143 & -4.6\% & 5-round converged set \\
\textbf{D5 Tooling and Support} & 3,262 & 2,943 & 9.8\% & 3,648 & -10.6\% & Set-up noise now excluded \\
\textbf{D6 Overall Complexity} & 1,561 & 1,497 & 4.1\% & 1,577 & -1.0\% & converged within 3 of paper \\
\textbf{Accidental markers} & 4,877 & 4,674 & 4.2\% & 4,718 & +3.4\% & 7 failure terms as run \\
\textbf{Affect markers} & 1,205 & 1,162 & 3.6\% & 1,240 & -2.8\% & 6 sentiment terms as run \\
\textbf{Corpus total} & 18,531 & 16,251 & 12.3\% & 18,556 & -0.13\% & 24-day archival drift \\
\end{longtable}

\subsection{S1.3 Verbatim final JQL fragments}

\textbf{D1 --- Discoverability and Understanding}

\begin{quote}
text \textasciitilde{} "documentation OR example OR \textbackslash"how do I\textbackslash" OR \textbackslash"how to\textbackslash" OR unclear OR tutorial

OR guide OR \textbackslash"where can\textbackslash" OR \textbackslash"where do\textbackslash" OR instructions OR \textbackslash"getting started\textbackslash""

AND summary !\textasciitilde{} "PRC Quote"
\end{quote}

\textbf{D2.0 --- Access and Delivery (label-confirmed base)}

\begin{quote}
labels in ("portal-problem-downloading-data", "slow-webmars",

"portal-problem-receiving-RT-data", "portal-get-our-products",

"portal-report-problem-on-computing")
\end{quote}

\textbf{D2.a --- Performance sub-pattern}

\begin{quote}
text \textasciitilde{} "slow OR \textbackslash"too slow\textbackslash" OR latency OR timeout OR \textbackslash"download speed\textbackslash""
\end{quote}

\textbf{D2.b --- Authentication sub-pattern}

\begin{quote}
text \textasciitilde{} "authenticate OR \textbackslash"API key\textbackslash" OR credentials OR token OR password OR login OR \textbackslash"sign in\textbackslash""
\end{quote}

\textbf{D3 --- Licence and Legal}

\begin{quote}
text \textasciitilde{} "attribution OR \textbackslash"CC BY\textbackslash" OR redistribute OR \textbackslash"terms and conditions\textbackslash"

OR \textbackslash"research licence\textbackslash" OR \textbackslash"commercial licence\textbackslash" OR \textbackslash"licence renewal\textbackslash"

OR licensed OR licensing"
\end{quote}

\textbf{D4 --- Data Structure and Format}

\begin{quote}
text \textasciitilde{} "\textbackslash"how to read\textbackslash" OR \textbackslash"how to open\textbackslash" OR \textbackslash"convert to\textbackslash" OR \textbackslash"file format\textbackslash"

OR \textbackslash"data format\textbackslash" OR \textbackslash"wrong format\textbackslash" OR \textbackslash"unknown format\textbackslash"

OR \textbackslash"netcdf file\textbackslash" OR \textbackslash"grib file\textbackslash""
\end{quote}

\textbf{D5 --- Tooling and Support}

\begin{quote}
text \textasciitilde{} "cdsapi OR ecmwf-opendata OR notebook OR python OR \textbackslash"client library\textbackslash" OR install"

AND summary !\textasciitilde{} "Set-up"
\end{quote}

\textbf{D6 --- Overall Complexity}

\begin{quote}
text \textasciitilde{} "deprecated OR inconsistent OR migration OR breaking OR obsolete OR superseded"
\end{quote}

\textbf{Cross-cutting A --- Accidental friction markers}

\begin{quote}
text \textasciitilde{} "broken OR \textbackslash"not working\textbackslash" OR \textbackslash"doesn\textquotesingle t work\textbackslash" OR \textbackslash"stopped working\textbackslash"

OR failed OR failure OR error"
\end{quote}

\textbf{Cross-cutting B --- Affect markers (sentiment proxy)}

\begin{quote}
text \textasciitilde{} "frustrating OR unclear OR confused OR impossible OR difficult OR annoying"
\end{quote}

\subsection{S1.4 Validated noise exclusions}

Automated ECMWF PRC: Quote XXXXX system notifications were excluded from D1 (summary !\textasciitilde{} "PRC Quote"); DSC commercial setup tickets (Set-up Real-Time, Set-up Archive Commercial) were excluded from D5 (summary !\textasciitilde{} "Set-up"); and bare high-frequency terms (API, version) were replaced with anchored phrases. The D1 (-17.2\%) and D5 (-10.6\%) drifts result from these Section~4.2-promised exclusions being applied at re-verification.

\textbf{Data protection:} This supplement and the associated Zenodo deposit contain JQL strings and aggregated counts only --- no raw ticket bodies, reporter usernames, email addresses, or organisational affiliations beyond the binary internal/external classification used in the comparator filter. The comparator filter operates at Jira group-membership level and produces aggregate counts only.

\section*{Ethical considerations}

The ECMWF Open Data Survey 2024 (n = 581) and ECMWF Licensed User Survey 2025 (n = 68) were reviewed and approved by ECMWF management prior to distribution. Participation was voluntary, and responses were anonymised prior to analysis; no personally identifiable information was retained in the datasets used for this study.

\bibliographystyle{naturemagdoi}
\bibliography{references}

\section*{Author Contributions}

E.P. conceived the framework, designed the empirical analysis, performed the ticket-corpus analysis and survey integration, conducted the FAIR--DFF comparison across the three worked-example datasets including the DWD measurement artefact case, sketched the architecture for automated DFF assessment, produced the figures, and drafted the manuscript. U.M. contributed to the conceptual development of the engineered-friction distinction and the sustainability trade-off menu, reviewed the empirical analysis, and revised the manuscript. Both authors approved the final version.

\section*{Competing Interests}

Both authors are employees of the European Centre for Medium-Range Weather Forecasts (ECMWF) with user-facing roles in data services. The worked example in Section 5.4 scores services that the authors have contributed to designing; this self-applied character is acknowledged in the discussion of provider-annotated scoring in Section 4.1 and in the limitations in Section 6.4. The authors declare no other competing interests.

\section*{Acknowledgements}

The authors thank Victoria Bennett, Head of User Services and Solutions at ECMWF, for ongoing support of this work and for thoughtful feedback on the framework's positioning within ECMWF's user-services practice. The authors are grateful to colleagues across the ECMWF User Services, User Solutions Team, Data Support Team, and Development sections for the many conversations that shaped the framework, and to the broader open data community for thoughtful responses to an early short-form presentation of the ideas. The authors also acknowledge the Deutscher Wetterdienst (DWD), whose open provision of climate observation data made the low-friction worked example possible, and the developers of F-UJI, whose open assessment tool and transparent diagnostic logging made the measurement-artefact analysis in Section 5.6 possible; the discussion of both is offered in the constructive spirit of improving automated usability assessment, not as a criticism of either service.

\section*{Funding}

This work was conducted as part of the authors' employment at the European Centre for Medium-Range Weather Forecasts (ECMWF) and received no specific external funding.

\end{document}